# Enhanced ambient stability of exfoliated black phosphorus by passivation with nickel nanoparticles

Maria Caporali,[a*] Manuel Serrano-Ruiz,[a] Francesca Telesio,[b] Stefan Heun,[b] Alberto Verdini,[c] Albano Cossaro,[c] Matteo Dal Miglio,[d] Andrea Goldoni,[d] Maurizio Peruzzini[a*]

[a] CNR ICCOM, Via Madonna del Piano 10, 50019 Sesto Fiorentino, Italy.

[b] NEST, Istituto Nanoscienze-CNR and Scuola Normale Superiore, Piazza S. Silvestro 12, 56127 Pisa, Italy.

[c] TASC CNR IOM, Area Science Park, Basovizza, 34149 Trieste, Italy.

[d] Elettra-Sincrotrone Trieste, Area Science Park, S.S. 14 km 163.5, I-34149 Trieste, Italy.

**E-mail**: maria.caporali@iccom.cnr.it, mperuzzini@iccom.cnr.it

## Abstract

Since its discovery, the environmental instability of exfoliated black phosphorus (2D bP) has emerged as a challenge that hampers its wide application in chemistry, physics, and materials science. Many studies have been carried out to overcome this drawback. Here we show a relevant enhancement of ambient stability in few-layer bP decorated with nickel nanoparticles as compared to pristine bP. In detail, the behavior of the Ni-functionalized material exposed to ambient conditions in the dark is accurately studied by TEM (Transmission Electron Microscopy), Raman Spectroscopy, and high resolution X-ray Photoemission and Absorption Spectroscopy. These techniques provide a morphological and quantitative insight of the oxidation process taking place at the surface of the bP flakes. In the presence of Ni NPs, the decay time of 2D bP to phosphorus oxides is more than three time slower compared to pristine bP, demonstrating an improved structural stability within twenty months of observation.

## Introduction

Within the family of two-dimensional (2D) materials, exfoliated black phosphorus (2D bP) has emerged as a hot and highly intriguing research topic over the last few years [1]. This novel 2D material is endowed with a unique structure and outstanding physical properties, thus it has a broad prospect in the construction of electronic devices [2]. In this respect, a prominent feature of 2D bP is that it is a semiconductor with a thickness-dependent direct band gap, ranging from 0.3 eV (bulk material) to approximately 2.0 eV (monolayer, called phosphorene) [3, 4]. Besides, the high mobility of charge carriers makes it a solid candidate for nanoelectronics [5]. Given the presence of a lone electron pair on each P-atom available to chemically interact with electron-poor atoms or moelcules, 2D bP suffers from a higher chemical reactivity in comparison to other 2D materials, which implies a lower stability in ambient conditions, being reactive with water and oxygen, and its reactivity is strongly accelerated by light [6-8]. Thus, exfoliated black phosphorus needs to be protected by a capping layer [9-14] in order to avoid the formation of various phosphorus oxides species, which inevitably cause surface degradation. On the other hand, the reactivity of few-layer bP can be exploited for various functionalizations, with molecules [15, 16], metal adatoms [17] or metal nanoparticles [18-20]. In particular, functionalization of bP with metal nanoparticles provided novel hybrids active in catalysis [18-20], electrocatalysis (*i.e.* hydrogen evolution reaction) [21], Li storage [21] and in biomedicine [22].

## Materials and Methods

*Functionalization of bP with nickel nanoparticles*

Bulk black phosphorus (bP) [23] and colloidal nickel nanoparticles [24] were prepared as previously described in literature. Liquid phase exfoliation of the bulk material was carried out under the action of ultrasound by using dimethylsulfoxide as solvent in inert atmosphere following a procedure optimized in our laboratories [25]. The synthesis of the nanohybrid Ni/2D bP was performed according to our published procedure [26]. To a freshly prepared suspension of few-layer bP in tetrahydrofuran (1.2 mL, 1.2 mg, 0.0387 mmol P) a black colloidal suspension of nickel nanoparticles dispersed in toluene (205 µL, 0.018 M, 0.00368 mmol, P/Ni = 10.5) was added drop-wise under nitrogen at room temperature. After stirring for 25 minutes, degassed acetone (5.0 mL) was added, and the mixture was centrifuged at 8000 RPM for 15 minutes. The black residue was washed once more with acetone (5.0 mL) and dried under a stream of nitrogen. The Ni content of the isolated nanohybrid was measured by ICP-MS (Inductively Coupled Plasma-Mass Spectrometry) and resulted equal to 15.2 wt.%. The average size of Ni NPs is (11.9±0.8) nm, see Figure S1.

The NiNPs functionalized bP solution and a solution of pristine bP were drop-casted on different substrates, suitable for the different characterization techniques. The samples of pristine 2D bP and Ni/2D bP were aged storing them under ambient conditions. In detail, TEM samples were kept at T = (23 ± 3)°C, in the dark with a relative humidity h = (40 ± 10)%, see Figure S2. Raman measurements of the samples were made in another laboratory and the Raman samples were kept at T = (23.4 ± 1.5)°C and at relative humidity h = (30 ± 8)%, see Figure S3.

*Characterization*

A transmission electron microscope (Philips) was used to obtain images of the materials. The incident energy of the electron beam was set to 80 kV. Dispersed fresh suspensions of 2D bP and Ni/2D bP in tetrahydrofuran (0.1 mg bP/mL in both cases) were drop-casted on carbon coated TEM grids and dried under a stream of nitrogen.

Raman spectroscopy was carried out on the same 2D bP and Ni/2D bP samples used above for TEM. For this purpose, the suspensions in tetrahydrofuran were drop-casted on a $SiO_2$/Si wafer, kept for few minutes, then rinsed with ethanol and dried under a stream of nitrogen. A Renishaw InVia micro-Raman was used, equipped with an optical microscope and a motorized stage. A laser power of 0.546 mW was used, with a spot size of 2 µm diameter, to avoid any photon-induced damage to the sample. High resolution X-ray Photoemission and Absorption Spectroscopy was performed at the Aloisa beamline [27] at the Elettra Synchrotron Radiation facility in Trieste (Italy), using suspensions of 2D bP and Ni/2D bP in tetrahydrofuran (3 mg bP/mL in both cases) that were drop-casted on Au(111)/mica, kept for 10 minutes, rinsed with ethanol, and dried under a stream of nitrogen.

**Results and discussion**

**TEM study**

Both on the pristine 2D bP sample and on the Ni/2D bP sample stored under ambient conditions in the dark, several flakes were chosen for TEM inspection and measured as prepared and after regular time intervals, to look for signatures of degradation. Figure 1 shows a TEM series for a pristine bP flake. Most of the pristine bP flakes were partially oxidized and hydrolyzed after one week, and within 15 days they lost completely their initial morphology and appeared as drops, see Figure 1 and Figure S4. The proposed degradation mechanism [28] goes via the leading role of water that being highly polar, interacts with oxygen adsorbed on the surface of bP and facilitates the electron transfer from bP to oxygen thus starting the oxidation process. Once P-O and P=O bonds are formed, being the P-oxides highly hygroscopic, they react with ambient water yielding molecular compounds, *i.e.* phosphorus oxyacids, $H_3PO_2$, $H_3PO_3$ and $H_3PO_4$. This explains why the flake morphology is lost, and only drops are present.

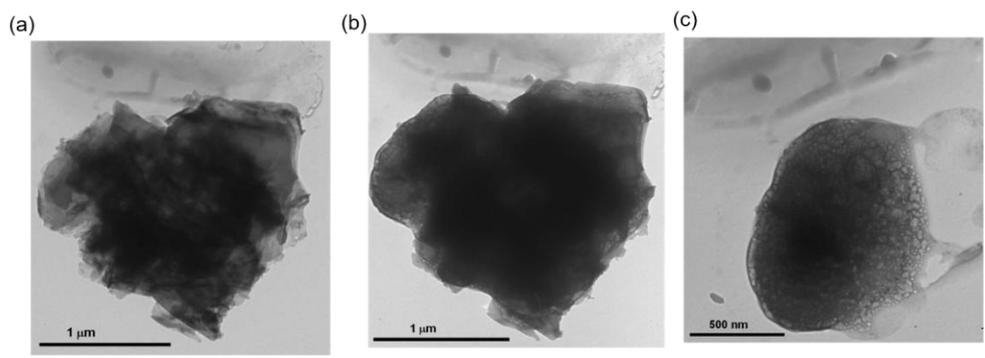

**Figure 1**. TEM images of a pristine BP flake: a) freshly prepared, b) after one week, c) after two weeks in ambient conditions. All images show the same flake.

Due to the nature of bP, Ni/2D bP flakes are also expected to react with oxygen and water, but surprisingly, after one week of ambient exposure, the flakes were completely preserved. Figure 2 shows the topography of the evolution of a selected flake. With prolonged time of observation, the flake appears covered by bubbles (Figures 2c-2f) due to the formation of P-oxides and P-oxyacids. Nevertheless, the morphology is preserved intact during seven months of aging, as shown in Figure 2, panels (c)-(f), suggesting that the oxidation process involves only the surface layers without going in depth.

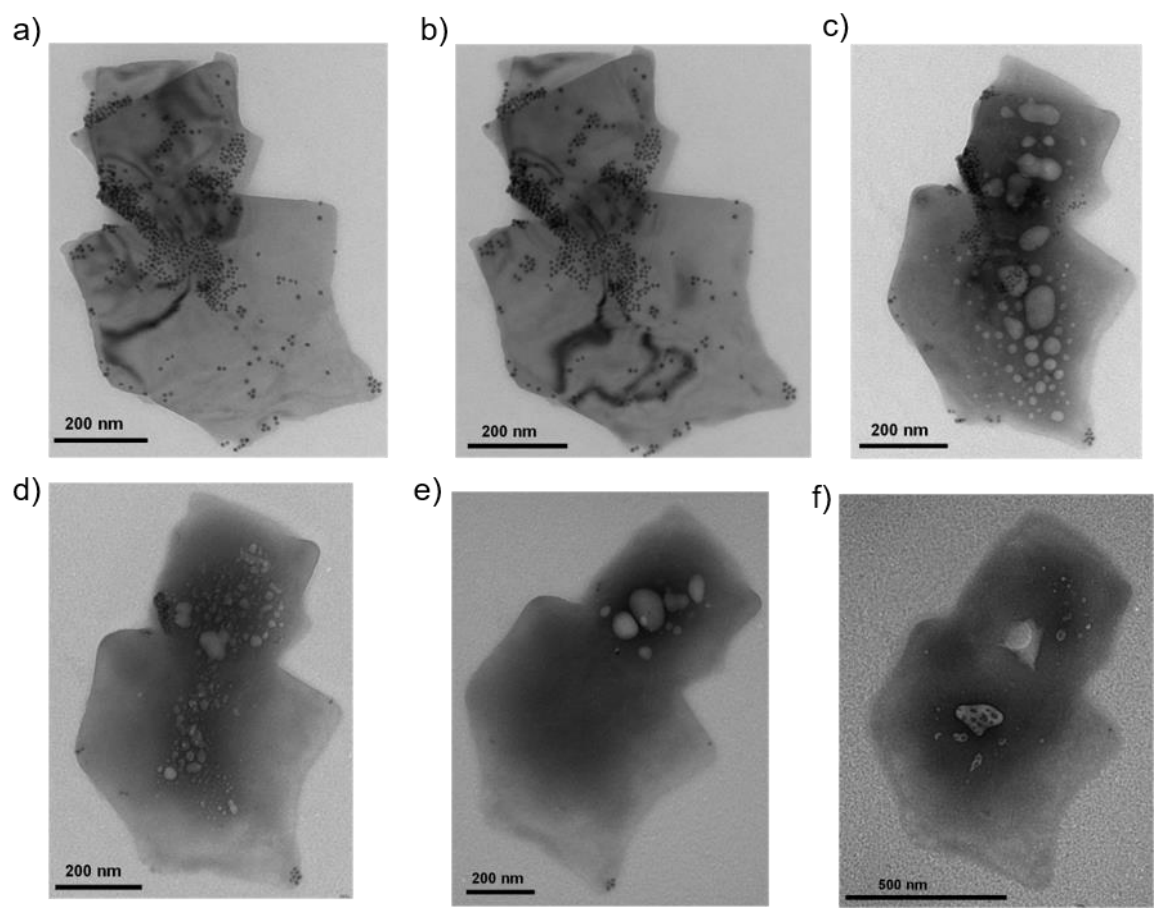

**Figure 2**. TEM images of a Ni/2D bP flake: a) freshly prepared; b) after one week; c) after 18 days; d) after one month; e) after three months, f) after seven months. The aging is performed keeping the samples under ambient conditions in the dark. All images show the same flake.

The lateral dimensions, length and width, of the observed flake (named afterwards flake 1) were obtained from TEM images and are plotted in Fig. 3. It clearly appears that the flake keeps roughly its size for two months, which indicates that the presence of the nickel nanoparticles stabilized it during this period. Afterwards there is a dramatic increase in size, which is then approximately constant during the subsequent three months of observation. In other words, the degraded area increases fast between the second and third month of aging, and afterwards it reaches a saturation, following an S-shaped growth curve. A similar behavior has been observed for bP flakes completely coated with 5 nm thick $Al_2O_3$ and exposed to ambient conditions [29]. It is remarkable that in our case the surface coverage with Ni NPs ranges from 8% to 15% (as calculated from TEM images) far away from being a thick layer covering 100% of the surface of the bP flake.

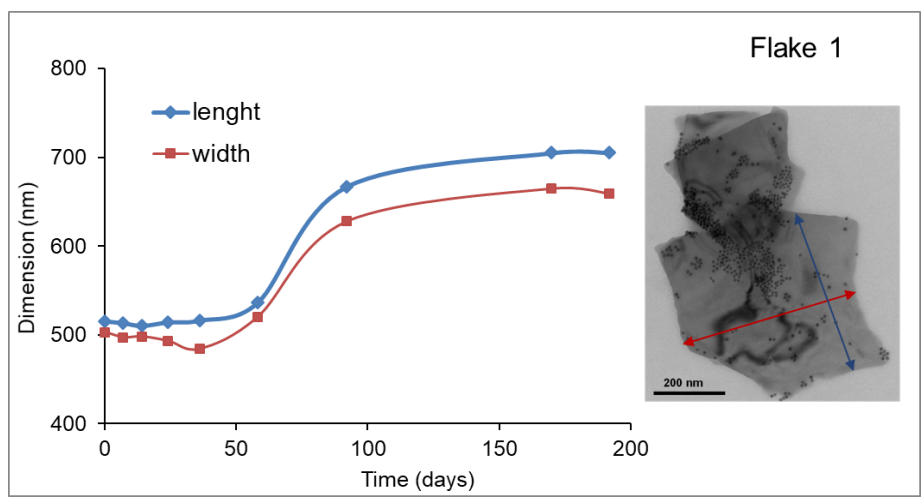

**Figure 3**. Variation of the lateral dimensions of the Ni/2D bP flake, length (blue) and width (red) during six months of ambient exposure.

Other Ni/2D bP flakes were studied by TEM, as well, within the six months of ambient exposure. All of them preserved their morphology. Their lateral dimensions were found to be enlarged up to 60 % in comparison to the fresh sample, following the same trend discussed above for flake 1, see Table 1 and Figures S5-S7.

**Table 1**. Variation of the lateral dimensions (in nm) of Ni/2D bP flakes kept in ambient conditions.

| Entry | dimension | day 0 | 2 months | Δ% 2 months | 3 months | Δ% 3 months | 6 months | Δ% 6 months |
|---|---|---|---|---|---|---|---|---|
| Flake 1 | lenght | 515 | 536 | 4.1 | 667 | 29.5 | 705 | 36.9 |
| | width | 503 | 520 | 3.4 | 628 | 24.9 | 659 | 31.0 |
| Flake 2 | lenght | 1107 | 1040 | -6.1 | 1290 | 16.5 | 1307 | 18.1 |
| | width | 768 | 758 | -1.3 | 1000 | 30.2 | 972 | 26.6 |
| Flake 3 | lenght | 704 | 742 | 5.4 | 910 | 29.3 | 970 | 37.8 |
| | width | 165 | 166 | 0.6 | 260 | 57.6 | 265 | 60.6 |
| Flake 4 | lenght | 737 | 760 | 3.1 | 1000 | 35.7 | 1012 | 37.3 |
| | width | 381 | 394 | 3.4 | 539 | 41.5 | 550 | 44.4 |

A comparison between the crystallographic structure of 2D bP [30] and $P_2O_5$ (which is also a layered material featuring an in-plane anisotropic structure) [31] shows a remarkable expansion of the unit cell going from pristine bP to the oxidized species: the size of the unit cell in the zig-zag direction increases from 3.316(1) Å to 4.890(4) Å ($\Delta = 47.5\%$), and in the armchair direction from 4.379(1) Å to 7.162(7) Å ($\Delta = 63.5\%$), as shown in Figure S8. Thus, the observed increase of the lateral dimensions of the flakes is an indication that $O_2$ reacted with bP forming P-O bonds. As will be discussed later, the presence of phosphorus oxides is further confirmed by x-ray photoelectron spectroscopy (XPS) measurements.

**Raman study**

The Raman signature of bP is very clear and consists of three main peaks [6] at 361 cm$^{-1}$, 438 cm$^{-1}$, and 466 cm$^{-1}$. They are labeled as $A^1_g$, $B_{2g}$, and $A^2_g$, and they are related to the out-of-plane, in-plane along zigzag, and in-plane along armchair vibrational modes, respectively [32]. Since these vibrational modes are strongly suppressed by oxidation [6], Raman spectroscopy is another powerful tool to study bP aging. These measurements were performed on dropcasted samples. Because of drop-casting, the small 2D bP flakes (both pristine and functionalized) tend to aggregate during solvent evaporation: a first morphological information that we can extract from the freshly prepared samples is that aggregation is much stronger in the case of pristine bP, with respect to the functionalized bP. In Figure 4 three typical areas for each sample are shown.

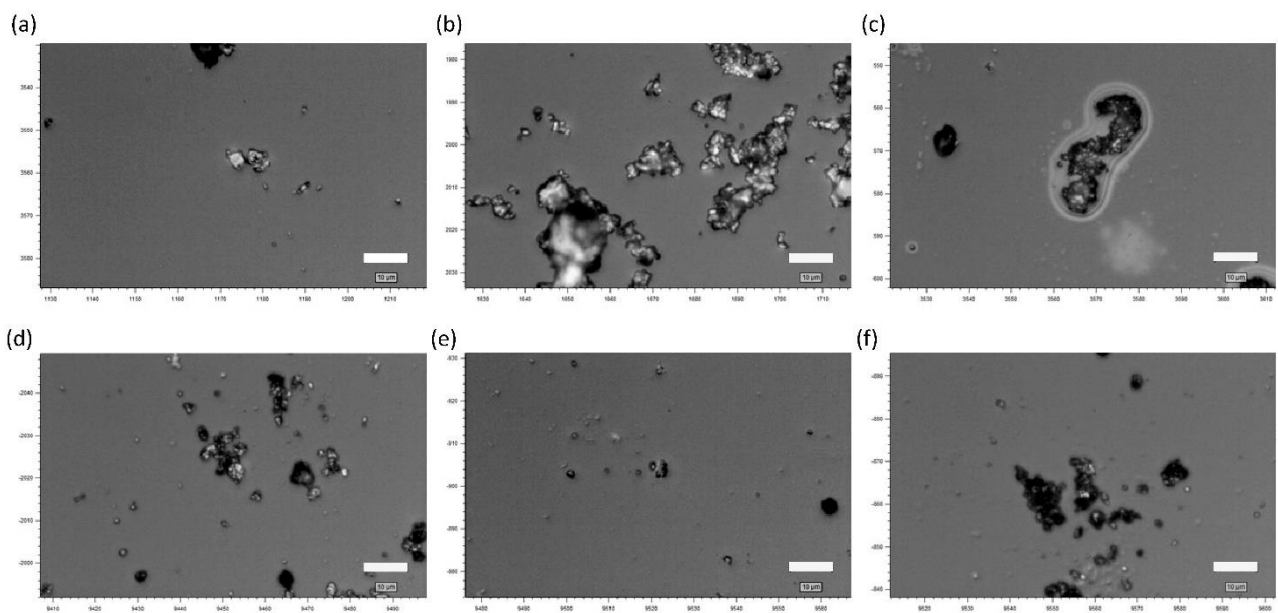

**Figure 4**. Typical morphology of aggregates of a freshly prepared 2D bP sample [(a) – (c)] and of a Ni NPs functionalized 2D bP sample [(d) – (f)]. Aggregation is more pronounced for the pristine sample. Scale bar = 10 µm.

Since aggregates of various morphology and size are present on the substrate, ranging from less than one µm to several 10 µm in lateral size, a statistical analysis of the Raman data has been performed, measuring several flakes and aggregates of different morphology – up to 45 at a time – to capture as much as possible the aging trend of the overall sample. Over all the experiment, more than 200 aggregates have been measured for each of the two samples.

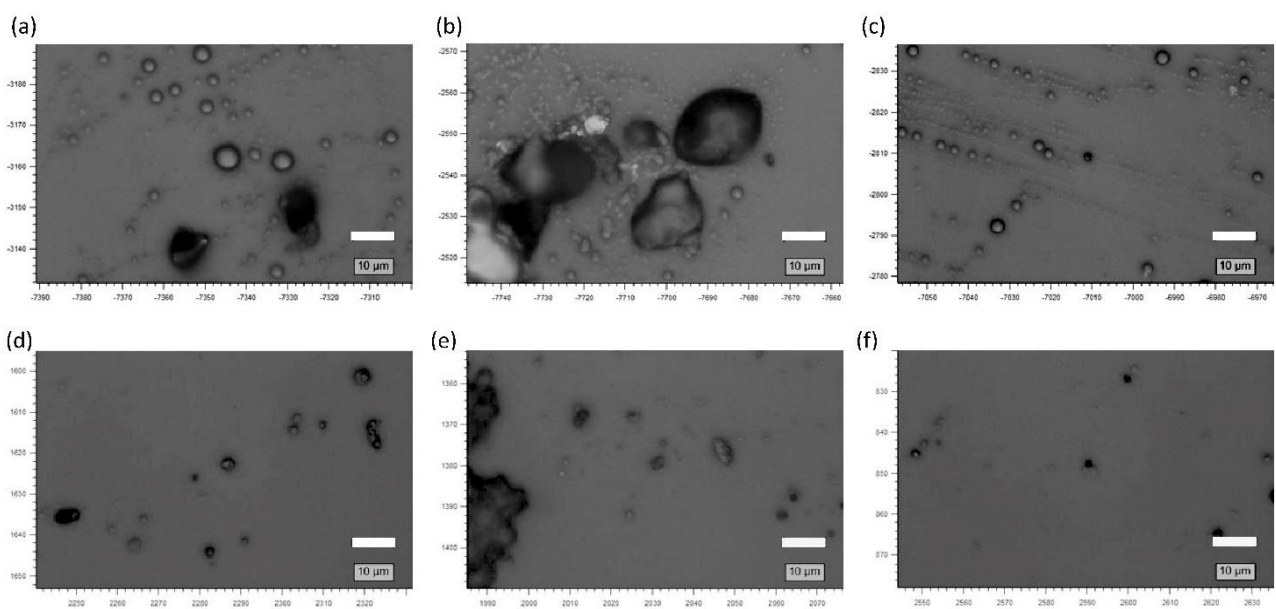

**Figure 5**. Typical morphology of the pristine sample - panels (a)-(c) - and of the Ni NPs functionalized sample - panels (d)-(f) - after 15 months of aging. A more pronounced hydrolysis is present on the pristine 2D bP sample, while even small aggregates display a quite clear structure for the Ni NPs functionalized sample. Scale bar = 10 µm.

The morphology at the end of month 15 is shown in Figure 5. While the pristine sample [images (a)–(c)] shows severe marks of degradation, the functionalized sample [images (d)–(f)] shows still several small aggregates which are just slightly degraded.

The averaged Raman spectrum from all aggregates measured at the beginning of the experiment, normalized to the Si peak intensity of the substrate, is presented in Figure 6(a) for the pristine bP sample and in Fig. (b) for the functionalized one. The Si peak at 520 cm$^{-1}$ is more pronounced for the functionalized sample because the aggregates have a much smaller lateral size, often smaller than the diameter of the laser spot. Averaged normalized spectra after 15 months of exposure to ambient conditions are presented in Figure 6(c) for the not-functionalized sample and in Figure 6(d) for the functionalized one. The relative decrease of the bP peaks intensity is much stronger for the non-functionalized sample, for which many of the measured aggregates were not Raman active anymore, while the Ni NPs functionalized sample displays still all Raman peaks with higher intensity.

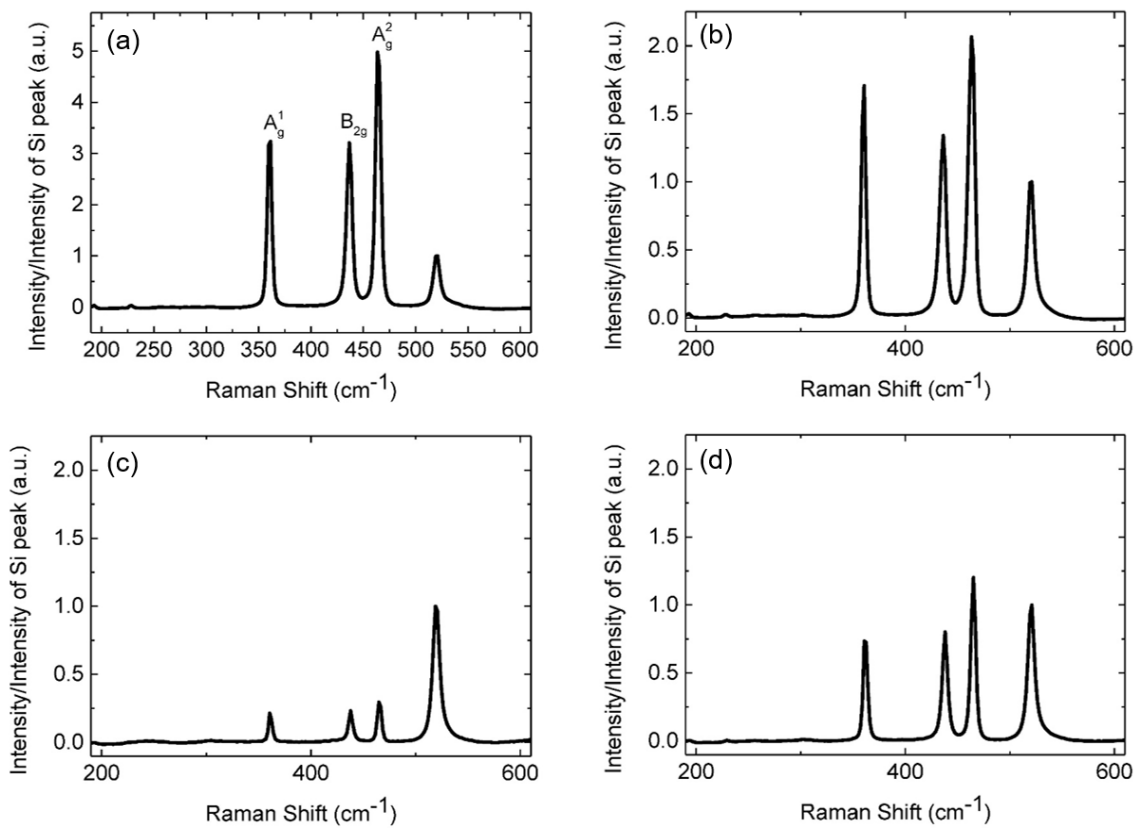

**Figure 6**. Averaged Raman spectra for (a) the pristine bP sample and (b) the Ni/2D bP sample at the beginning of the experiment. (c) and (d), the same after 15 months of exposure to ambient conditions.

All spectra were acquired in the same conditions. Spectra for the same sample are averaged, then the background is subtracted, and finally the spectra are normalized to the Si substrate peak intensity.

To visualize the trend of the Raman activity with time, we present the aggregate Raman data as a function of time in Figure 7. Since the intensity of the bP peaks is not just related to aging, but as already discussed also to the initial thickness and lateral size of the aggregates, as a first indicator for the good preservation of the sample, we use the Raman activity in itself – namely the presence of the bP peaks in the Raman spectra, independent of their intensity. For each date, the number of Raman active aggregates with respect to the total number of measured aggregates is reported, attributing as an error bar the standard error of the distribution, evaluated as $1/\sqrt{N}$, where N is the total number of measurements. Degradation is more pronounced for the pristine bP sample, but given the large variety in the aggregates' morphology, the two samples are still compatible within the error bars even after 12 months. However, in the last part of the experiment, and in particular for the last measurements performed after 20 months, the two distributions are well separated. This clearly shows that the Ni/2D bP sample degrades much more slowly than the pristine one. This result is even more striking when this information is complemented with the morphological information. From the beginning of the experiment, the Ni/2D bP aggregates are much smaller, as visible from the optical microscopy images in Fig. 4. In general, larger and thicker bP flakes degrade more slowly, because of their lower surface-to-volume ratio and consequently their lower reactivity [6]. Here, however, we observe that the larger pristine bP aggregates degrade much faster than the smaller Ni/2D bP aggregates. This is an unequivocal proof of the enhanced environmental stability of the Ni/2D bP sample. At the end of the experiment, just $(22 \pm 15)\%$ of the aggregates belonging to the non-functionalized sample were still Raman active, while of the Ni NPs-functionalized aggregates were $(73 \pm 15)\%$.

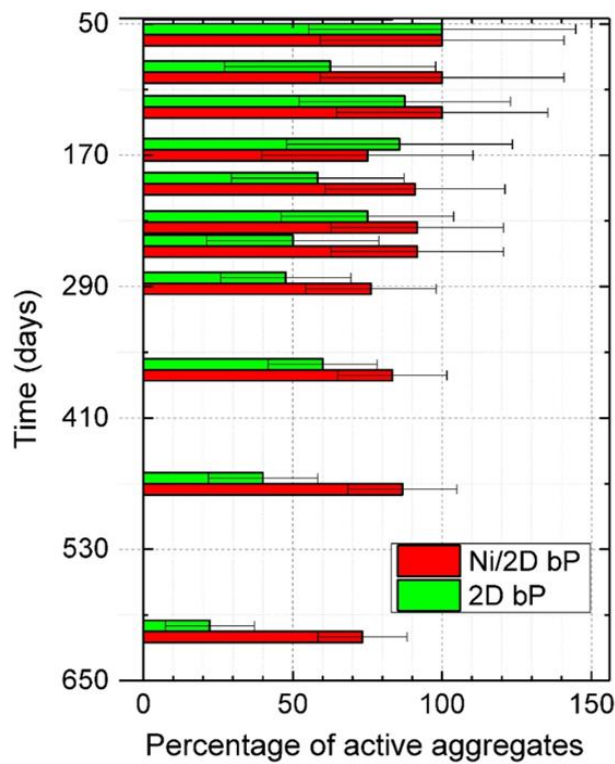

**Figure 7**. Number of Raman active aggregates for the not functionalized sample (green bars) and for the functionalized sample (red bars). Measurements before day 45 are not shown to enhance the clarity of the graph, since they are all compatible within the error bar with 100% of Raman activity.

To quantify the decay in Raman activity, a plot of the $A^2_g$ peak intensity, normalized to the intensity of the Si peak, is presented in Figure 8 as a function of time for (a) the pristine sample and (b) the functionalized one. This analysis has been done with the average spectra (with 5 days binning), to overcome as much as possible the scattering due to the morphological differences among flakes.

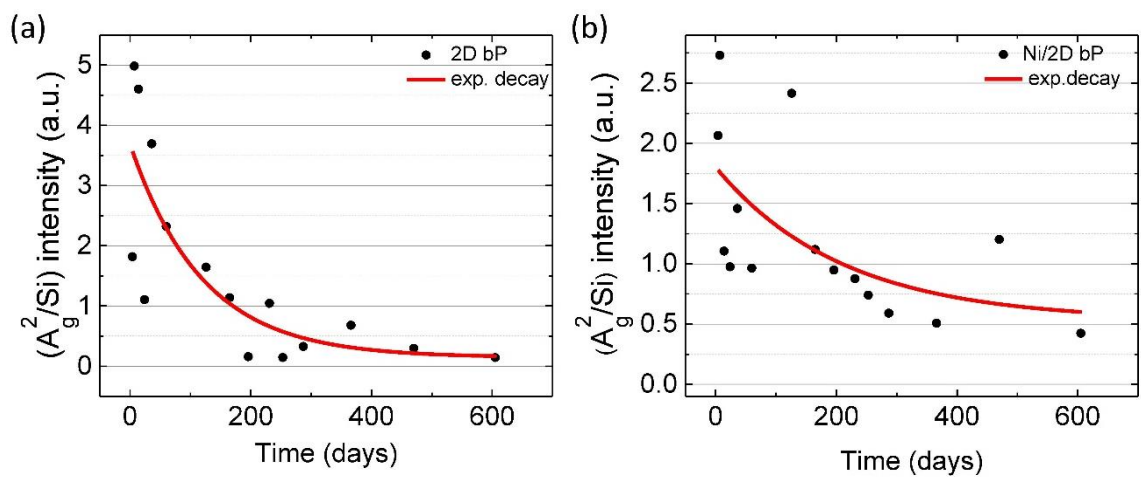

**Figure 8**. Averaged intensity of the $A^2_g$ peak normalized to the intensity of the Si peak, plotted as a function of time. The individual spectra are grouped with a 5 days binning and averaged. Then the background is subtracted and the intensity of the $A^2_g$ peak is normalized to the Si peak. The trend is fitted with an exponential decay.

The smaller intensity of the normalized $A^2_g$ peak for the functionalized sample with respect to the pristine at the beginning of the experiment is consistent with the data presented in Figure 6.

The normalized $A^2_g$ intensity data as a function of time have been fitted with an exponential decay $\propto e^{-t/\tau}$, where $t$ is time and $\tau$ is decay time, as previously reported by Favron *et al.*[6]. The exponential decay captures well the trend of the data, as shown in Figure 8. We obtain a decay time $\tau \approx 90\ days$ for the non-functionalized bP sample and $\tau \approx 300\ days$ for the one functionalized with NPs, confirming the faster degradation of the non-functionalized sample. The decay for the Ni NPs-functionalized sample is therefore more than three times slower than for pristine bP, in agreement with what observed in Figure 6. This clearly proofs that Ni NPs slow down the degradation of exfoliated bP.

**X-ray Photoemission and Absorption Spectroscopy study**

To study quantitatively the role played by the presence of Ni NPs on the environmental stability of bP and to evaluate the amount and the identity of oxidized species formed on the surface, XPS were measured on a fresh sample of Ni/2D bP and after aging in ambient conditions for 18 days. In parallel, a sample of pristine bP kept in the same conditions was examined as a reference. The binding energy scale has been calibrated by measuring the Au $4f_{7/2}$ peak taken as reference [33]. As shown in Figure 9a, the XPS spectrum of a sample of fresh 2D bP exhibits two maxima at 130.0 and 130.85 eV, which are characteristic of the spin-orbit split phosphorus P 2p peak positions, namely P $2p_{3/2}$ and P $2p_{1/2}$, respectively [34]. The deconvolution of the XPS data was carried out using a fitting function with one spin-orbit doublet (P1) plus up to 3 Gaussian functions (P2, P3, P4) and the corresponding Shirley background [34, 35]. We name P1 the peak of not oxidized phosphorus, P4 the peak corresponding to phosphorous in its highest oxidation state (+5), and in some cases we added also two more peaks, P2 and P3, representing P-oxides with intermediate oxidation states +1 and +3. Figure 9b shows the XPS spectrum of a fresh sample of Ni/2D bP, which has the same features as pristine bP, plus a broad peak at 134.2 eV attributed to an incipient formation of phosphorus oxide species. Indeed, the presence of electronegative groups such as oxygen or hydroxide on the surface of bP withdraws charge leading to more tightly bound phosphorus electrons. As a result, $P_xO_y$ species experience a higher binding energy for P 2p core electrons in comparison to un-oxidized material [36].

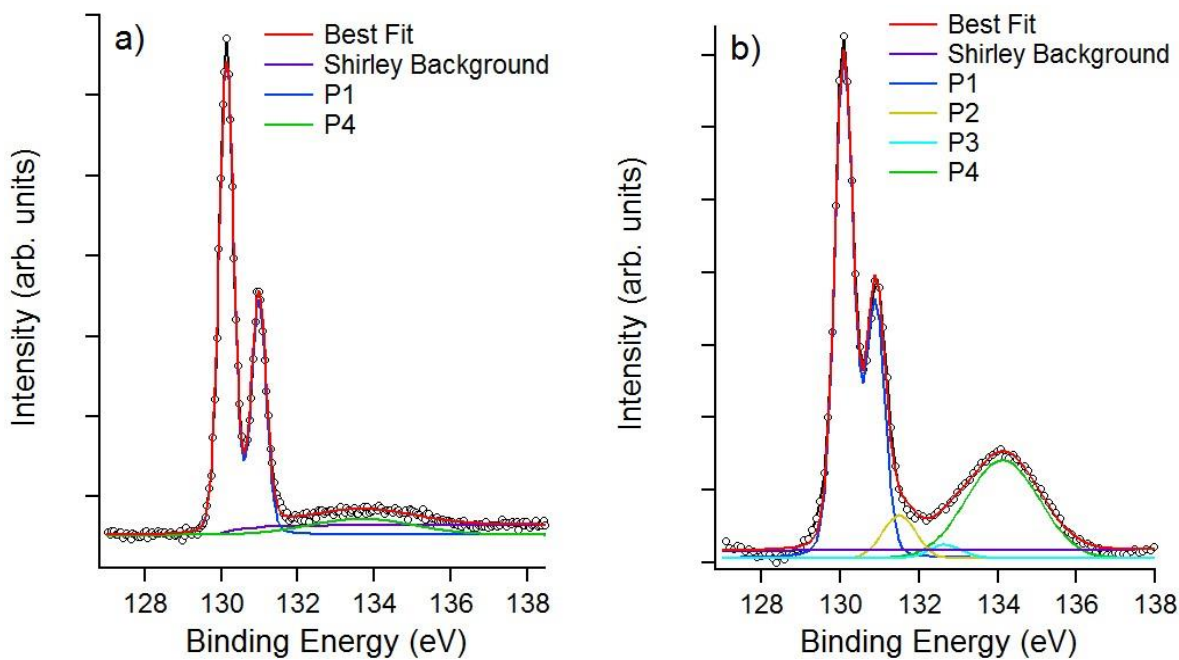

**Figure 9**. P 2p core level XPS spectra. a) XPS spectrum of a fresh sample of 2D bP dropcasted on Au/mica, b) XPS spectrum of a fresh Ni/2D bP sample on Au/mica. Both spectra were taken at h$\nu$ = 415 eV.

Figure 10 shows a comparison between XPS spectra of non-functionalized bP (left side) and Ni/2D bP (right side) both aged in a dark environment for the same time (18 days). Intriguingly, while the peaks related to P-P bonds remain at the same binding energy (B.E.) (130.0 and 130.85 eV, respectively, for the P 2p doublet) in both samples, in the case of P-O bonds, the peak maximum is observed at different binding energies: 134.1 eV in Ni/2D bP and 134.9 eV in pristine bP, indicating different compositions of phosphorus oxide species. On the basis of calculations by D. Tománek *et al*. [37], 2D bP oxides with oxidation state of +1, +3, and +5 have core-level binding energies in the range of 132-136 eV. Thus, a shift of 0.8 eV to higher B.E. suggests a larger amount of P-oxides having oxidation state +5, *i.e.* phosphorus pentoxide, $P_2O_5$. Consequently, a higher degree of oxidation in the case of pristine bP with respect to Ni/2D bP is confirmed. To evaluate the thickness of the oxide layer, XPS measurements were carried out on both samples using three different energies for the incident photons: 415 eV, 515 eV, and 1000 eV, in order to increase progressively the photoelectron mean free path and thus the sampling depth of the experiment. Increasing the photon energy from 415 eV to 1000 eV, there is a gradual decrease of the intensity of the peak corresponding to P-O bonds. This is a clear signature that only a thin oxide layer is present atop of the bP flakes.

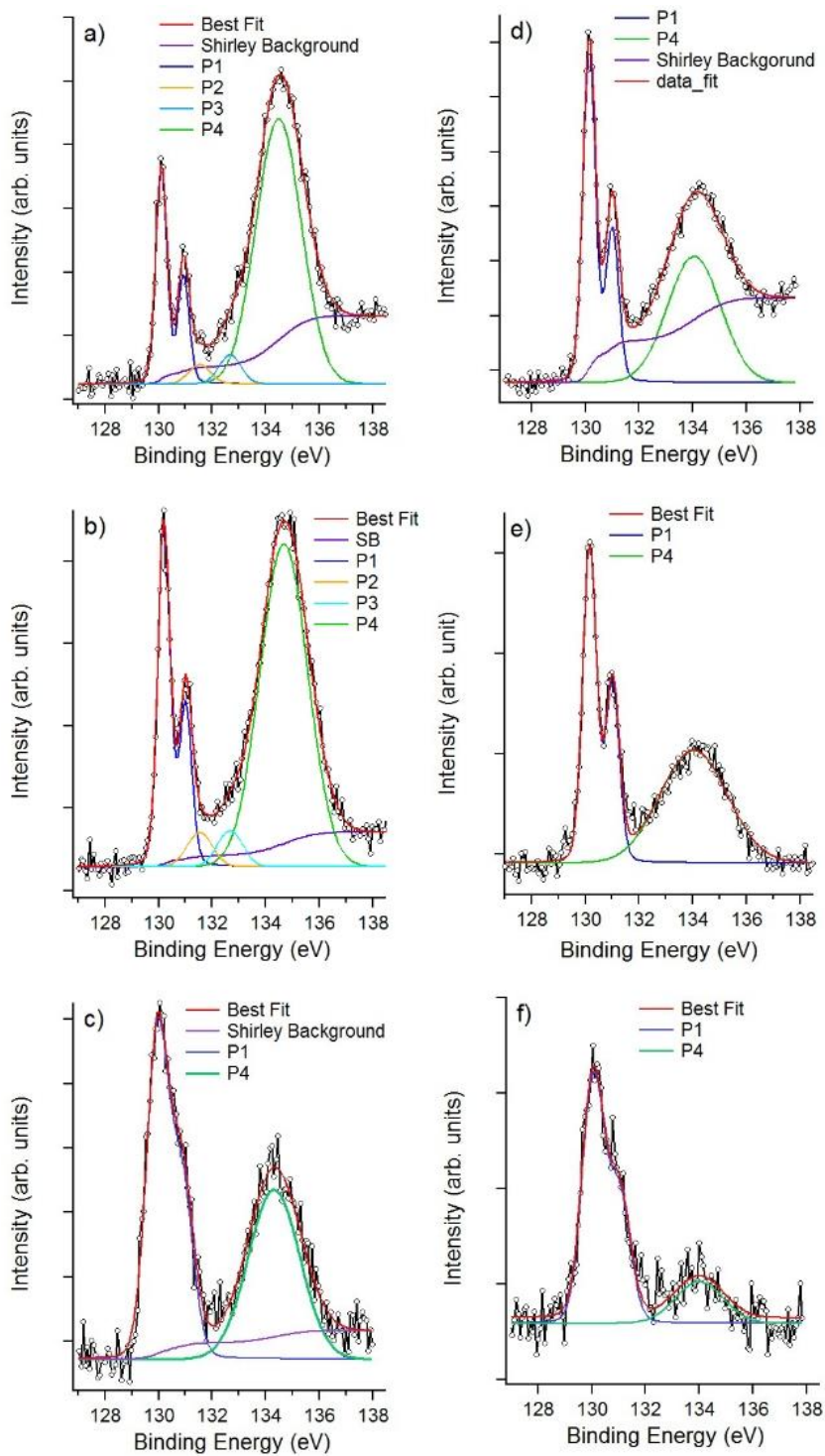

**Figure 10**. Depth-dependent photoelectron spectroscopy of the P 2p core level of 2D bP (left side, panels a-c) and Ni/2D bP (right side, panels d-f) after 18 days of air exposure, taken at hν = 415 eV (a, d), 515 eV (b, e) and 1000 eV (c, f).

The thickness of the phosphorus oxide layer was estimated according to the equation: $d = \lambda \ln[(I_{oxide}/I_{bP})+1]$ where $\lambda$ is the photoelectron inelastic mean free path [38], and $I_{oxide}$ and $I_{bP}$ are the area in the XPS spectra under the component of P-oxides and elemental phosphorus, respectively.

Thus, we determine an average thickness of phosphorus oxide of $(13.9 \pm 0.4)$ Å for pristine bP and of $(6.2 \pm 0.4)$ Å for Ni/2D bP after aging. Being the thickness of a single layer of planar $P_2O_5$ equal to 5.5 Å according to literature [39, 31] this means that we have three layers of $P_2O_5$ on top of pristine bP and approximately one layer on top of Ni/2D bP, confirming that the corrosion caused by oxygen and water is much deeper in the case of pristine bP. A further confirmation of the protective role of Ni NPs comes from the comparison of the ratio between P1 and P4 areas as a function of the inelastic mean free path shown in Fig S9, where the signal corresponding to not oxidized P is dramatically damped down when Ni NPs are missing.

In the region where the XPS signature of Ni is expected, no significant signal could be acquired, because of the relatively low amount of Ni with respect to bP (see Experimental Section). Therefore, NEXAFS (Near Edge X-Ray Absorption Fine Structure) measurements were carried out, because this technique has a higher sensitivity. Figure S10 shows the NEXAFS spectrum collected on a sample of Ni/2D bP freshly prepared. The characteristic $L_3$ and $L_2$ peaks of metallic nickel at 852.7 eV and 870.0 eV, respectively, are detected while the peak at 854.0 eV due to Ni(II) is absent, which means that nickel nanoparticles are pure Ni(0), and there is no oxidation. Running NEXAFS on the aged sample of Ni/2D bP, we detected again the characteristic $L_3$ signal of metallic nickel at 852.7 eV plus a shoulder at ~ 854.5 eV, suggesting the presence of Ni(II) due to the formation of a NiO layer as outer shell of the nanoparticles, being a consequence of the aging in ambient conditions, see Figure S11.

The different degradation rate in Ni/2D bP with respect to 2D bP can be undoubtedly related to the presence of Ni NPs, although they only cover roughly from 8% to 15% of the nanosheets' surface. Thus, we cannot ascribe to Ni NPs the role of a physical barrier against oxidation, we can invoke a role, mainly of chemical nature, played by Ni NPs which are expected to interact with defect sites of 2D bP surface, such as steps and edges, being the more reactive. In this way, they prevent those sites to be approached by oxygen and water molecules. The overall degradation process results markedly slowed down, meaning that also the reactivity of bare P atoms, i.e. not directly interacting with Ni NPs, is slowed down. This can be interpreted on the basis of our recent results [20] on 2D bP functionalized with Pd NPs, where a coordination bond of covalent nature was found between P atoms and outer shell of Pd NPs. On this ground, a covalent bond between Ni NPs and 2D bP is also expected, and since it is known [41] that covalent modifications of 2D bP may induce variations of electronic properties, and shifts of CBM (conduction band minimum) below the redox potential of $O_2\text{-}O_2^-$, than it is easily explained the passivation observed in our functionalized bP nanosheets.

## Conclusions

The simultaneous influence of two factors, air and ambient humidity (up to 50%), has been evaluated in the stability of 2D bP functionalized with Ni NPs. The three techniques used, TEM, Raman, and high resolution XPS, are in very good agreement in showing an increased chemical stability of Ni/2D bP flakes, that maintain intact their size for one month. The presence of nickel nanoparticles on the surface of 2D bP makes the rate of the oxidation process more than three times slower as derived from the decay of $A^2_g$ Raman signal, and a thinner layer of phosphorus pentoxide is formed as measured by XPS. Remarkably, in our experiments the surface coverage of bP flakes by Ni NPs is only partial, thus we cannot invoke a physical barrier of protection made by the metal nanoparticles but their chemical interaction with bP should be addressed. Thus, a large part of the bP surface is uncovered and remains available for further functionalization and applications in devices that are necessarily kept in ambient conditions and in which access to the surface is crucial, such as gas sensing.

## Acknowledgement

Thanks are expressed to EC for funding the project PHOSFUN "*Phosphorene functionalization: a new platform for advanced multifunctional materials*" (ERC ADVANCED GRANT N. 670173 to M.P.). F.T. thanks CNR-NANO for funding the SEED project SURPHOS (*Surface properties of few layer black phosphorus investigated by scanning tunneling microscopy.*) S.H. acknowledges support from Scuola Normale Superiore, project SNS16_B_HEUN – 004155. Mr. Carlo Bartoli (CNR ICCOM) is acknowledged for his skillfull technical assistance. A. V., A.G. and M.D.M. are partly supported by the NATO project G5140.

## Supporting information

Histogram relative to the size distribution of Ni NPs is given in S1. Graphs related to T (°C) and humidity to which samples were exposed along six months are shown in S2-S3. TEM images of a reference pristine bP flake is given in S4, graphs of the variation of the lateral dimensions versus time of selected Ni/2D bP flakes are given in S5-S7, crystallographic structures of bP and $P_2O_5$ are given in S8, the P1/P4 ratio versus photon energy in aged samples is given in S9, NEXAFS of Ni/2D bP are given in S10-S11.

## ORCID IDs

**Caporali M** http://orcid.org/0000.0001-6994-7313

**Peruzzini M** http://orcid.org/0000.0002-2708-3964

**Notes**

The authors declare no competing financial interests.

# Supplementary Information to "Enhanced ambient stability of exfoliated black phosphorus by passivation with nickel nanoparticles"

Maria Caporali,[a*] Manuel Serrano-Ruiz,[a] Francesca Telesio,[b] Stefan Heun,[b] Alberto Verdini,[c] Albano Cossaro,[c] Matteo Dal Miglio,[d] Andrea Goldoni,[d] Maurizio Peruzzini[a*]

[a] CNR ICCOM, Via Madonna del Piano 10, 50019 Sesto Fiorentino, Italy.

[b] NEST, Istituto Nanoscienze-CNR and Scuola Normale Superiore, Piazza S. Silvestro 12, 56127 Pisa, Italy.

[c] TASC CNR IOM, Area Science Park, Basovizza, 34149 Trieste, Italy.

[d] Elettra-Sincrotrone Trieste, Area Science Park, S.S. 14 km 163.5, I-34149 Trieste, Italy.

**E-mail**: maria.caporali@iccom.cnr.it, mperuzzini@iccom.cnr.it

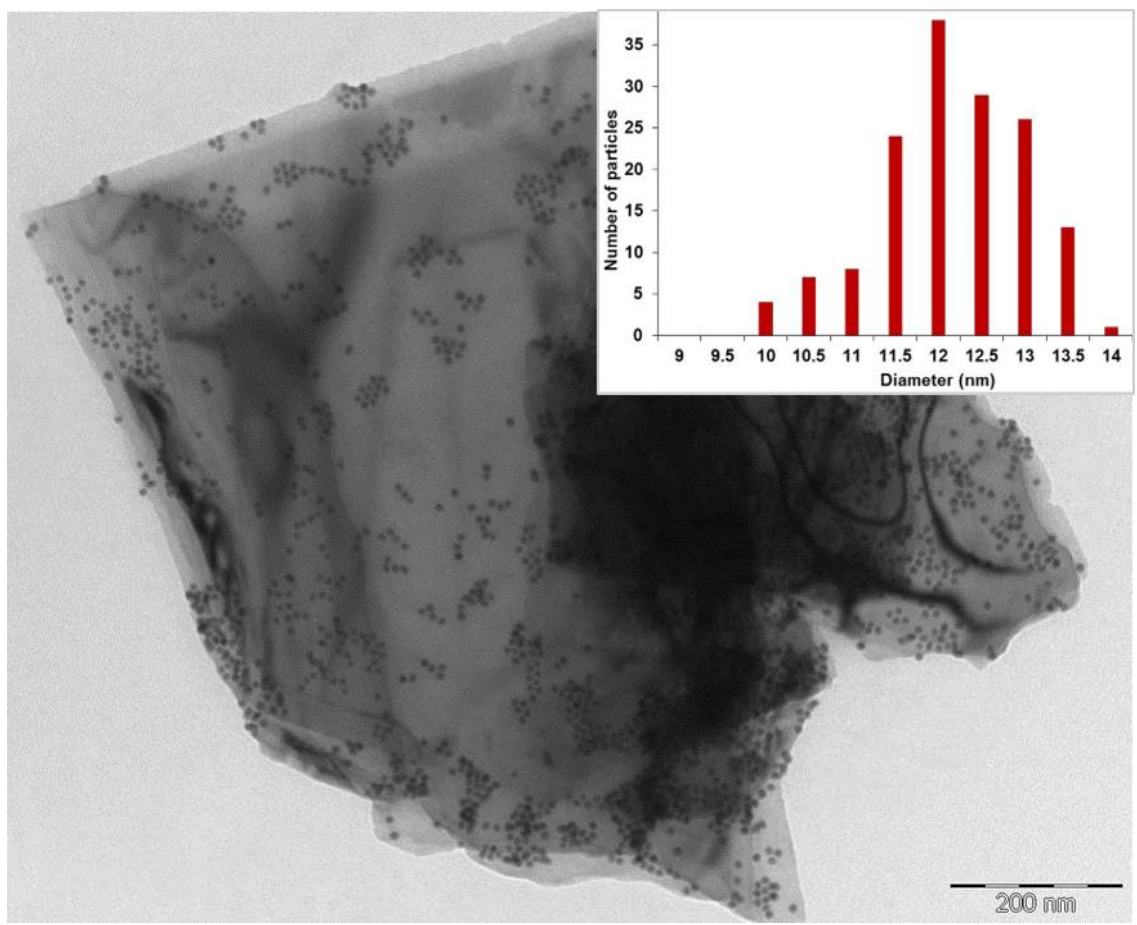

Figure S1. Bright field TEM image of Ni/2D bP. Scale bar: 200 nm. The inset shows the relative size distribution histogram of the Ni NPs. The average size of the Ni NPs is (11.9 ±0.8) nm.

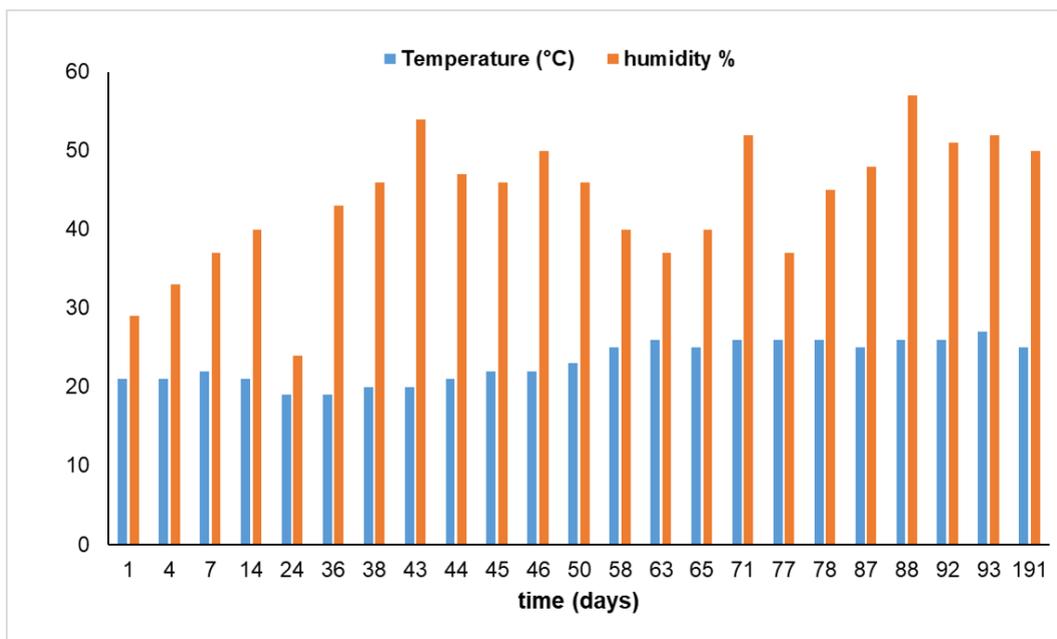

Figure S2. Ambient conditions, temperature (T) and humidity (h), to which TEM samples were exposed for six months. Average T = (23±3) °C, average h = (40±10)%.

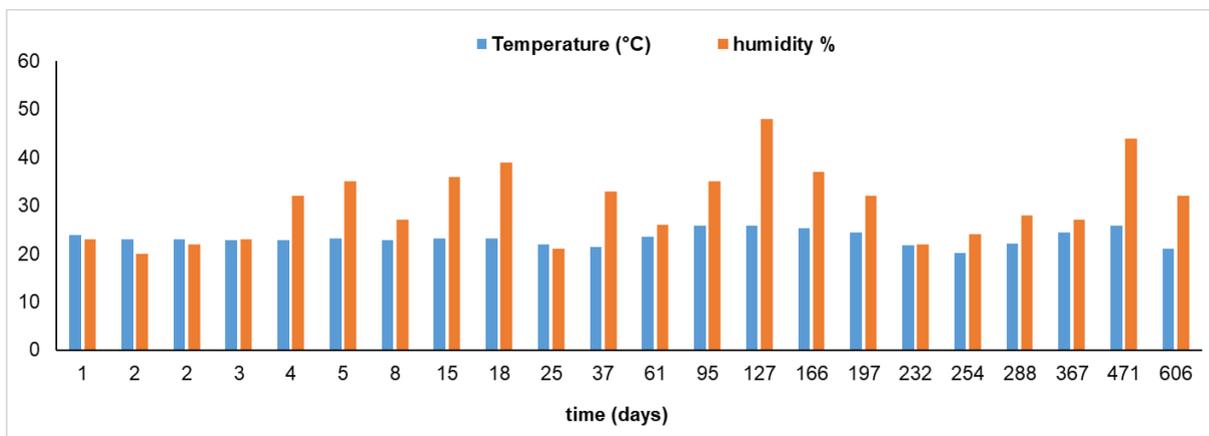

Figure S3. Ambient conditions, temperature (T) and humidity (h), to which Raman samples were exposed for twenty months. Average T = (23±1.5) °C, average h = (30±8)%.

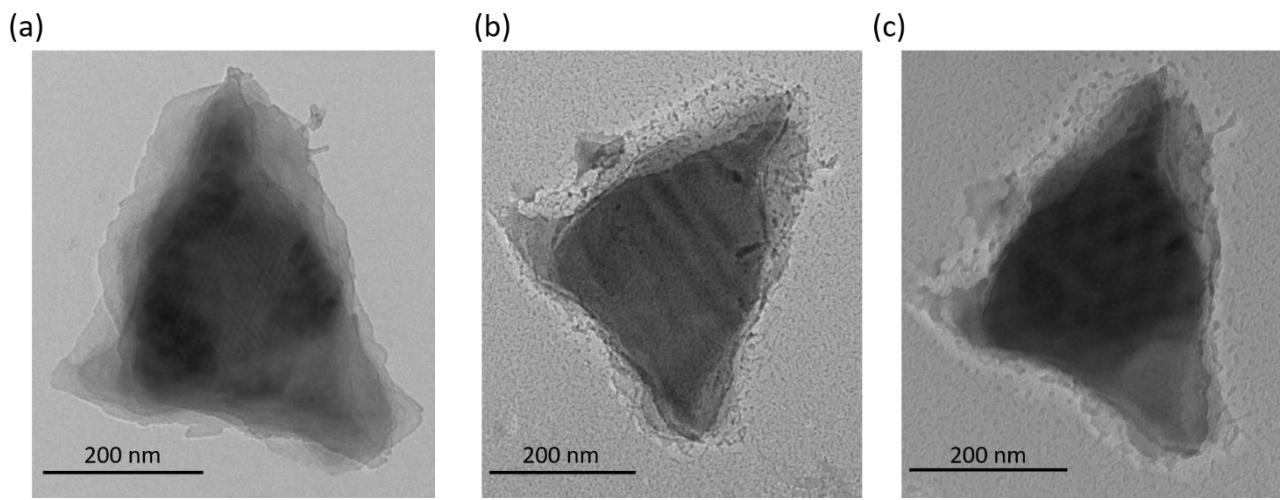

Figure S4. TEM images of a pristine bP flake: a) freshly prepared, b) after two weeks, c) after two months under ambient conditions.

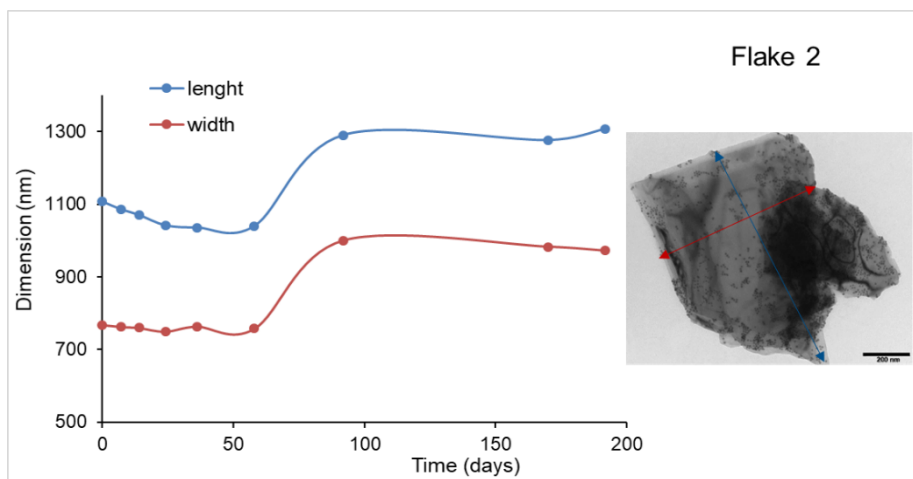

| day | lenght (nm) | width (nm) |
|-----|-------------|------------|
| 0   | 1107        | 768        |
| 7   | 1086        | 763        |
| 14  | 1070        | 760        |
| 24  | 1042        | 750        |
| 36  | 1036        | 764        |
| 58  | 1040        | 758        |
| 92  | 1290        | 1000       |
| 170 | 1276        | 983        |
| 192 | 1307        | 972        |

Figure S5. Variation of the lateral dimensions of flake 2, during six months of ambient exposure.

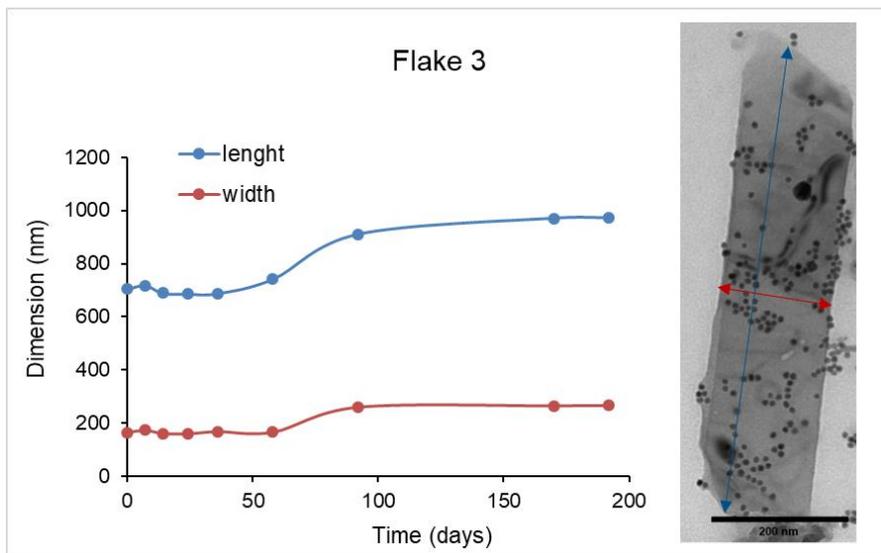

| day | lenght (nm) | width (nm) |
|-----|-------------|------------|
| 0 | 704 | 165 |
| 7 | 718 | 173 |
| 14 | 689 | 161 |
| 24 | 686 | 160 |
| 36 | 688 | 168 |
| 58 | 742 | 166 |
| 92 | 910 | 260 |
| 170 | 970 | 265 |
| 192 | 971 | 266 |

Figure S6. Variation of the lateral dimensions of flake 3, during six months of ambient exposure.

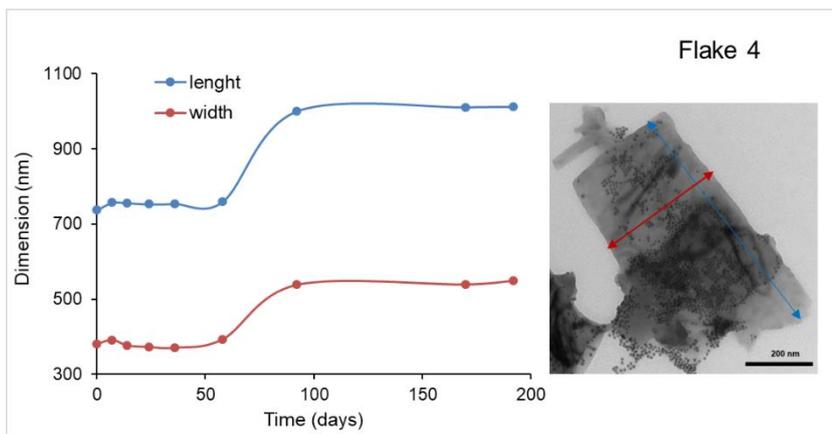

| day | lenght (nm) | width (nm) |
|-----|-------------|------------|
| 0 | 737 | 381 |
| 7 | 757 | 392 |
| 14 | 756 | 377 |
| 24 | 753 | 373 |
| 36 | 754 | 371 |
| 58 | 760 | 394 |
| 92 | 1000 | 539 |
| 170 | 1011 | 540 |
| 192 | 1012 | 550 |

Figure S7. Variation of the lateral dimensions of flake 4 during six months of ambient exposure.

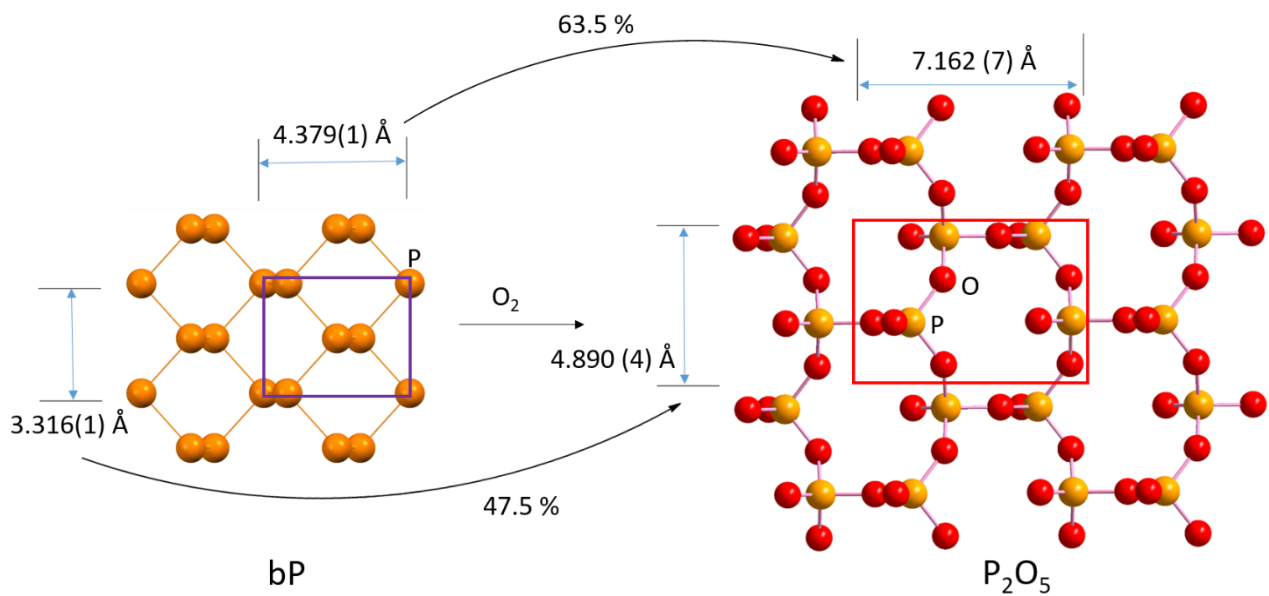

Figure S8. Comparison of the crystallographic structure of bP (left) and P$_2$O$_5$ (right).

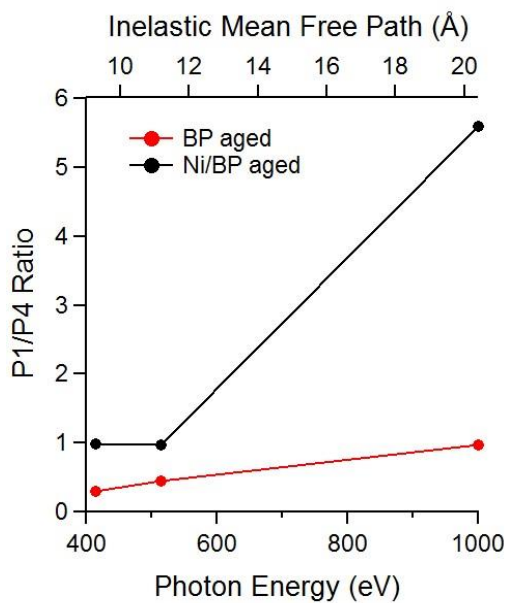

Figure S9. The ratio between the areas underneath the peaks labeled P1 (related to P-P bonds) and P4 (related to P-oxides) is reported as a function of the photon energy (bottom axis) and for the calculated photoelectron inelastic mean free path (top axis).[1]

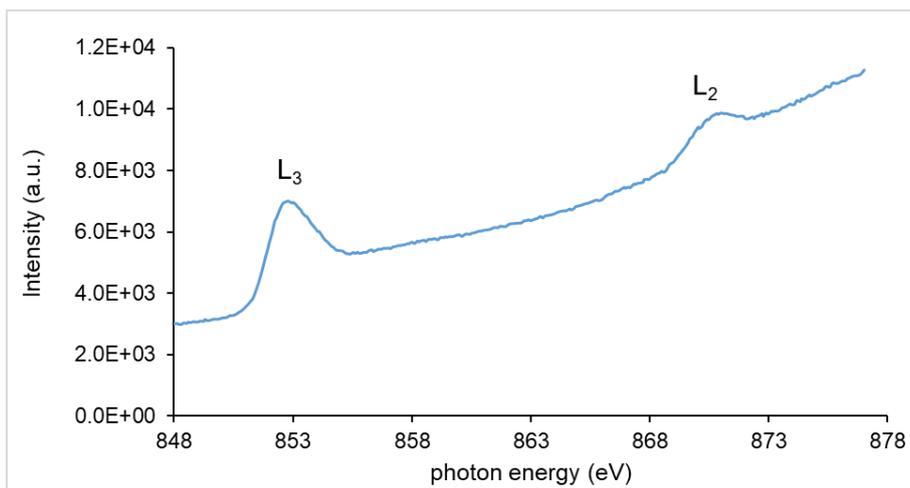

Figure S10. NEXAFS of L$_{2,3}$ edge of nickel in a fresh sample of Ni/2D bP.

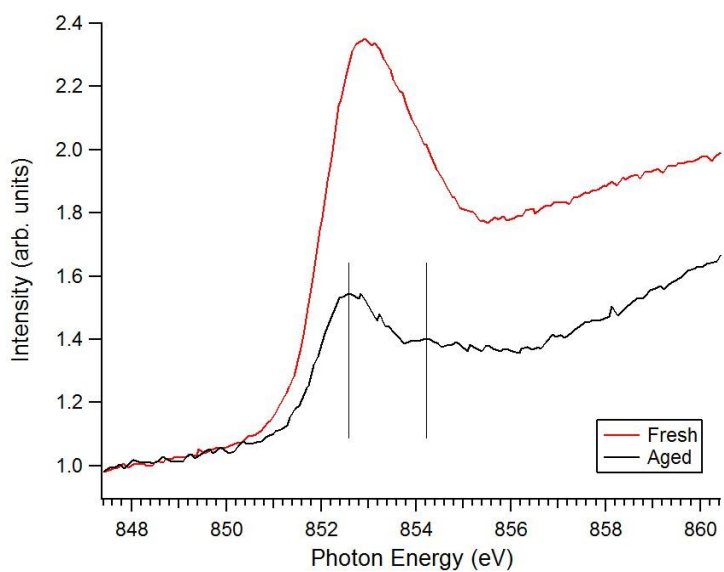

Figure S11. NEXAFS of the L$_3$ edge of nickel. a) Red line: fresh sample of Ni/2D bP; black line: Ni/bP aged 18 days.